\documentclass{JHEP} 
\usepackage{epsfig}
 \newcommand{\IR}{{\bf R}}

\def\ket#1{\left| #1\right\rangle}

\newcommand{\be}{\begin{equation}}
\newcommand{\ee}{\end{equation}}

\preprint{MIT-CTP-3097\\ \\ {\tt hep-th/0103103}}

\title{Testing the Uniqueness of the Open Bosonic String Field Theory Vacuum}
\author{Bo Feng, Yang-Hui He and Nicolas Moeller 
\footnote{
Research supported in part by
the CTP and LNS of MIT and the U.S. Department of Energy 
under cooperative research agreement \# DE-FC02-94ER40818.
N. M. and Y.-H. H. are also supported by the Presidential Fellowship
of MIT.}
\\
Center for Theoretical Physics,
\\ Massachusetts Institute of Technology,\\ Cambridge, MA 02139, USA.\\
\email{fengb, yhe, moeller@ctp.mit.edu}
}
\abstract{The operators $K_n$ are generators of 
reparameterization 
symmetries of Witten's cubic open string field theory.
One pertinent question is whether they can be utilised to generate
deformations of the tachyon vacuum and thereby violate its
uniqueness. We use level truncation to show  that these transformations 
on the vacuum are in fact pure gauge transformations 
to a very high accuracy, thus
giving new evidence for the uniqueness of the perturbatively stable
vacuum. 
Equivalently, this result implies the vanishing of 
some  discrete cohomology classes 
of the BRST operator in the stable vacuum.}
\begin{document}
\newpage
\section{Introduction}
In the last two years, there has been a huge amount of work done 
to understand tachyon condensation by using Witten's cubic open bosonic 
string field theory \cite{WITTENSFT}. 
The fate of a space-filling D25-brane in the open bosonic 
string theory is described by Sen's three conjectures 
(\cite{9902105}, \cite{9904207}). The first proposes that 
the difference in energy of the tachyon between the perturbative vacuum and 
the perturbatively stable vacuum exactly cancels the tension of the D-brane. 
The second asserts that after the tachyon condenses, all open string 
degrees of freedom disappear, leaving us with the closed string 
vacuum. The last conjecture states that non-trivial field configurations 
correspond to lower-dimensional D-branes.

The first and third conjectures have been shown to be true to 
a very high level of
accuracy (\cite{KS} - \cite{0008127}); 
they have also been proven
analytically in Boundary String Field Theory
(\cite{9208027} - \cite{0009191}).
The second conjecture however, is by now the most puzzling. Roughly,
it can be regarded at three different levels of stringency. A weak
statement is that all perturbative conventional open string 
excitations disappear from
the  perturbatively stable vacuum. There has been several works testing this
statement from various approaches:
one could show that some flat directions are removed,
as was done in \cite{0008033,0007153},
or that the kinetic terms of the  string field fluctuations are absent as in
\cite{0101162}, or by the usage of toy models in field theory
(\cite{0003278}, \cite{0008227}, \cite{0008231}, \cite{0009246}, 
\cite{0011226}, \cite{0102071}) as well as the boundary state formalism
(\cite{0103056, ShatashviliTalk}). 

A slightly stronger statement is that not only the conventional 
perturbative
open string excitations disappear, 
but more precisely the full cohomology 
of the new BRST operator around the tachyon vacuum vanishes. As usual,
the cohomology could include discrete states 
in addition to conventional excitations.
In \cite{0012251,0102112}, Rastelli, Sen and Zwiebach
have proposed that
after a field redefinition, the new BRST operator may be taken to be simply
$c_0$, or more generally a linear combination of operators
of the form $(c_n + (-)^n c_{-n})$. The cohomology of such operators
is manifestly trivial, and thus these authors are proposing this
more stringent form of the second conjecture.
Using this simple BRST operator on the vacuum, 
they were able to find solutions
corresponding to the D-25 brane and lower dimensional D-branes.

Finally, a third level of understanding the second conjecture is that
the perturbatively stable vacuum should correspond precisely to the
closed string vacuum. 
A possible interpretation of this statement is that we should
be able to isolate closed string excitations. Indeed, it is well-known
that closed string perturbative amplitudes can be in principle
isolated from cubic open string field theory diagrams. 
Thus closed string physics is there, though in a rather unmanageable form.
It may be that closed string states appear more manifestly
around the tachyon vacuum. If this is the case, perhaps one could
obtain a description which differs from the explicit one provided
by closed string field theory \cite{9206084}.
For recent discussions of closed strings in the tachyon vacuum
see \cite{0012081,0103056}.

A full understanding of Sen's conjectures, especially the second, would 
probably require the knowledge of the analytic solution for the 
perturbatively stable 
vacuum in cubic Open String Field Theory (OSFT), which is not yet known. However,
we can still make progress by using various methods; in particular we will
use the level truncation scheme 
to show that certain  deformations
from the perturbatively stable vacuum belong to the trivial cohomology 
of the BRST operator $Q_{\Psi_0}$ governing the spectrum of the 
string field theory around the tachyon vacuum. This provides
evidence for the second level of the second conjecture, viz., the disappearance
of discrete excitations. 

Our idea is the following. It is well known that the cubic OSFT has a {\em
reparameterization symmetry} generated by   operators (\cite{WITTENSSFT},
\cite{LPP}, \cite{gross-jevicki}, \cite{0006240}):
$$
K_n := L_n - (-)^n L_{-n}.
$$
Hence if $\Psi$ is a solution of the equation of motion, i.e.,
$Q_B \Psi+\Psi \star \Psi=0$, so is $e^{\epsilon K_n} \Psi$; this
follows immediately from properties (\ref{K1}-\ref{K2}) of 
$K_n$ which we will list in the next section.
In other words, we can generate new solutions by acting
$e^{\epsilon K_n}$ on a known solution, and in particular, the
perturbatively stable vacuum $\Psi_0$ of OSFT (which we will always
assume to lie in the Feynman-Siegel gauge).

A problem subsequently arises. From the physics point
of view, we expect the tachyon vacuum solution to be unique, i.e.,
there should be no moduli space of the tachyon vacuum solution. 
On the other hand we seem to be able 
to deform $\Psi_0$ by $e^{\epsilon K_n}$ with arbitrary 
parameters $\epsilon$ and $n$.

In order that this seeming paradox may be consistent with physical
intuition, there are two possibilities. 
Firstly it may be that 
$K_n\Psi_0=0$ for all $n$, which would imply that $e^{\epsilon
K_n}\Psi_0=\Psi_0$ and that no new tachyon vacuum solutions are generated.
At face value, this possibility
is very unlikely to be true because the action of $K_n$ takes 
a solution in the Siegel gauge out of it, and a miraculous cancellation
would be needed. In fact, we have verified that the $K_n$'s
do not annihilate the tachyon condensate. 
This leaves us with another choice, i.e., though
$\epsilon K_n \Psi_0$ may not vanish, it could be a {\em pure
gauge transformation} for any $n$ and $\epsilon$. 

The purpose of this note is to show that it is indeed the case that
$K_n \Psi_0$ is a pure gauge transformation.
Our result can be summarized as follows.
First by using a recursive relation obtained from the algebra of the
$K_n$'s, we show that
it is enough to demonstrate that if the action of $K_1$ and $K_2$ on
the tachyon vacuum $\Psi_0$ are pure gauge transformations, so too are
$K_n$ for all $n$. Then we use the level truncation scheme to
calculate $K_1 \Psi_0$ and $K_2\Psi_0$ up to levels 5 and 4 respectively. We
then show that they are indeed pure gauge transformations to an
excellent accuracy of $1.5\%$ for $K_1$ (resp. $1.6\%$ for $K_2$).

The statement that $K_n\Psi_0$ is a pure gauge transformation for any $n$
is equivalent to the assertion that the discrete zero momentum
state $K_n\Psi_0$ is $Q_{\Psi_0}$ exact. 
That is, these discrete BRST-closed states are actually BRST-trivial.
In a very nice recent work,  Ellwood and Taylor \cite{ellwood} 
have addressed the triviality
of the cohomology classes associated to continuous non-zero momentum
deformations of the tachyon vacuum. More precisely, they discuss the scalar
excitations at even levels and show that if they are $Q_{\Psi_0}$ closed, they are
$Q_{\Psi_0}$ exact also to very high accuracy, thus
giving the first convincing evidence for the disappearance
of (a subset) of the conventional open string excitations. Our results,
by focusing on discrete cohomology,  complement
their work.  Therefore, our
works jointly support, from different view-points,
the triviality of the cohomology and hence the validity of Sen's
second conjecture.

The outline of the paper is as follows. In Section 2 we review the key
properties of the $K_n$ operators and show that it suffices to
consider only $K_{1,2}$. Level truncation was subsequently applied in
Section 3 for $K_2 \Psi_0$ up to level 4, and in Section 4 for $K_1 \Psi_0$ up to
level 5 while most of the details of the involved computations are
left to the Appendix. Finally we end with concluding remarks and open
questions in Section 5.
\section{The $K_n$ Symmetry of Cubic String Field Theory}
It is a well known fact that the subalgebra\footnote{It is in fact the
maximal subalgebra that leaves the mid-point of the string invariant.}
of the Virasoro algebra
generated by the following operators
\begin{equation}
\label{K_n_sym}
K_n=L_n-(-)^n L_{-n},
\end{equation}
is a symmetry of Witten's Cubic String Field Theory
(\cite{WITTENSSFT,0012251}).  Because
$K_{-n} = (-1)^{n+1}K_n$ we need only consider the cases of $n \ge 1$.
These operators have the following properties:
\begin{eqnarray}
\label{K1} [K_n,Q_B] & = & 0  \\
\label{K2} K_n(A\star B) & = & (K_n A)\star B+ A\star (K_n B)  \\
\label{K3} \langle K_n A, B \rangle & = & - \langle A, K_n B \rangle \,,
\end{eqnarray}
where $A$ and $B$ are arbitrary string fields, and 
$Q_B$ is the conventional  BRST operator. Incidentally, comparing
(\ref{K2}) and (\ref{K3}) with similar properties  for $Q_B$, we notice that there
is no sign factor $(-1)^A$ here because 
$K_n$ is a ghost number zero Grassman even operator. 

Using (\ref{K2})
it is easy to show that $e^{K_n}(A \star B) = (e^{K_n} A) \star
(e^{K_n} B)$. Therefore if $Q_B \Psi + \Psi \star \Psi = 0$, so too is 
$Q_B (e^{K_n}\Psi) + (e^{K_n} \Psi) \star (e^{K_n}\Psi) = 0$, where we
have used (\ref{K1}). In other words, using the symmetry generators 
$K_{n}$,  
we can obtain new solutions of the equation
of motion by acting on a known solution. 
As we have argued in the
introduction, this poses a question about the uniqueness of the tachyon 
vacuum. On the one hand, from the physics point of view, we expect that the 
tachyon vacuum should be unique. On the other hand, we can seemingly generate new
solutions by acting $e^{K_n}$ on the vacuum. For these two ideas to
be consistent, we must propose that {\em the action of $K_n$ on the
tachyon vacuum $\Psi_0$ should be a pure gauge transformation}, i.e.,
\begin{equation}
\label{unique_tachyon}
K_n \Psi_0 \stackrel{?}{=}
 \delta \Psi_0 \equiv Q_{\Psi_0} \Lambda=
Q_B \Lambda +\Psi_0 \star \Lambda
-\Lambda \star \Psi_0.
\end{equation}
It is the checking of the conjecture (\ref{unique_tachyon}) with which
this present paper is concerned.
We remark in passing that there seems to be the possibility that $K_n
\ket{\Psi_0} = 0$. However this is highly unlikely because though
$\Psi_0$ is in the Feynman-Siegel gauge, the $K_n$ action does not
preserve this gauge. Indeed we have verified at low levels that this
triviality does not seem to be the case so that we need to return to
address (\ref{unique_tachyon}).

First we check the consistency of the conjecture. Because we have
$Q_{\Psi_0}Q_{\Psi_0}=0$ on the right hand side of
(\ref{unique_tachyon}) due to nilpotency,
so too must we get zero when we act
$Q_{\Psi_0}$ on the left. This is indeed so:
\begin{eqnarray*}
Q_{\Psi_0} K_n \Psi_0 & = & Q_B (K_n \Psi_0)+ \Psi_0 \star(K_n \Psi_0)
+(K_n \Psi_0)\star \Psi_0 \\
& = & K_n (Q_B  \Psi_0)+K_n (\Psi_0 \star \Psi_0)  \\
& = & K_n \{Q_B  \Psi_0+\Psi_0 \star \Psi_0\} \\
& = & 0,
\end{eqnarray*}
where in the second step we have used $[K_n,Q_B]=0$ (\ref{K1}) 
and in the last step, the equation of motion (the expression in the
braces) of $\Psi_0$.  
Notice that this check requires no usage of any
special properties of the tachyon vacuum, so for any solution of the
equation of 
motion $Q_B  \Psi+\Psi \star \Psi=0$, we always have $K_n \Psi$ being
$Q_{\Psi}$ closed.
Our conjecture is the statement that when
$\Psi=\Psi_0$ is the tachyon vacuum, $K_n \Psi_0$ is not only closed,
but also exact, whence BRST-cohomology trivial.
To show this is true is our work.

Naively it seems to be difficult to check that all the $K_n$ actions are 
mere pure gauge transformations because there are an infinite number 
of them. However, we can show that it suffices to check for
$K_1$ and $K_2$, then by iteration $n\geq 3$ follows.
This can be done in two steps.
Firstly we recall that the $K_n$'s form an algebra:
\begin{equation}
\label{K_relation}
	[K_n,K_m] = (n-m) K_{n+m} - (-1)^m (n+m) K_{n-m}.
\end{equation}
Secondly we can show that if for some $n$ and $m$,
$$
K_n \Psi_0 =Q_{\Psi_0} \Lambda_n,~~~~~K_m \Psi_0 =Q_{\Psi_0} \Lambda_m,
$$
then
\begin{eqnarray}
\label{K_nK_m}
[K_n,K_m] \Psi_0
& = & Q_B \tilde{\Lambda}_{n,m}+\Psi_0\star\tilde{\Lambda}_{n,m}-
\tilde{\Lambda}_{n,m}\star \Psi_0 = Q_{\Psi_0} \tilde{\Lambda}_{n,m},
\end{eqnarray}
and hence pure gauge, where 
\begin{equation}
\label{lambda_construction}
\tilde{\Lambda}_{n,m}= K_n \Lambda_m-K_m \Lambda_n+\Lambda_n\star 
\Lambda_m-\Lambda_m \star \Lambda_n.
\ee
Combining (\ref{K_relation}), (\ref{K_nK_m}) and 
(\ref{lambda_construction}), we see instantly that if the conjecture
is true for $K_1$ and $K_2$, then by iteration, we would have the
result for all $K_{n\geq 3}$.
\section{The Exactness of $K_2 \Psi_0$}
In this section, we check that $K_2 \Psi_0$ is a pure gauge
transformation, which would imply that $K_2 \Psi_0$ is BRST-exact.
First we do the calculation at level two, which is very simple. We use this
example to demonstrate our method, then we go further to level four. For the
details, the reader is referred to the Appendix.

Before proceeding, let us make some general 
remarks which is explained further in the Appendix.
The tachyon solution $\Psi_0$ of \cite{0002237} has only even level
components. So if the gauge parameter $\Lambda$ is in an even (resp. odd) level, 
$\Psi_0 \star \Lambda-\Lambda \star \Psi_0$ will contain only even (resp. odd)
levels as well; this is shown in (\ref{simplify}).
Furthermore, since
$Q_B$ does not change the level and $K_2$ increases or
decreases the level by two, to see whether $K_2$ on $\Psi_0$ is a pure gauge,
we can restrict the gauge parameters to be in even levels only. 
Likewise, for $K_1$, because it increases or
decreases the level by one, $K_1\Psi_0$ must have only odd levels.
Therefore, in this case we can restrict all gauge parameters to be in
odd levels only. In particular we will focus on levels $2,4$ for
$K_2$ and $3,5$ for $K_1$.
\subsection{Fitting at Level 2}
Up to level two, there are four components for the string field:
\begin{equation}
\label{tachyon_2}
\ket{\Psi}= \eta_{0,1}\ket{\Omega}+\eta_{2,1}b_{-1} c_{-1} \ket{\Omega}+
\eta_{2,2} b_{-2} c_{0} \ket{\Omega}+\eta_{2,3} L_{-2}^m\ket{\Omega},
\end{equation}
where the $\eta$'s are numerical coefficients and 
$L_{-n}^m$ are matter Virasoro operators. Furthermore,
$\ket{\Omega}=c_1\ket{0}$ and $\ket{0}$ is the $SL(2,\IR)$ invariant 
vacuum\footnote{Our
	notation is different from that in \cite{0002237}. We use
	here, for the matter part, the
	universal basis instead of the oscillator basis.}. 
For simplicity,
we denote the basis of the fields as a row vector with four
components so that
$$
(\eta_{0,1},\; \eta_{2,1},\; \eta_{2,2},\; \eta_{2,3}) 
:=
\eta_{0,1}\ket{\Omega}+\eta_{2,1}b_{-1} c_{-1} \ket{\Omega}+
\eta_{2,2} b_{-2} c_{0} \ket{\Omega}+\eta_{2,3} L_{-2}^m\ket{\Omega}.
$$
To this convention of notation of fields we shall adhere.

The numerical values for these coefficients have been computed to
great precision in the Feynman-Siegel gauge\cite{0002237}. 
At level $(2,6)$ (here we use their
convention that $(L,I)$ refers to truncating fields up to level $L$
and interactions up to level $I$; also we shall use their
normalization), the vacuum field (\ref{tachyon_2}) is
\begin{equation}
\label{level2_num}
(\eta_{0,1},\; \eta_{2,1},\; \eta_{2,2},\; \eta_{2,3})
=(0.39765,\; -0.13897, \; 0, \; 0.040893).
\end{equation}

Up to level two, for the gauge parameter $\ket{\Lambda}$ of ghost number
0, there is only one numerical parameter $\mu_{2,1}$:
\begin{equation}
\label{level2_gauge}
\ket{\Lambda}=\mu_{2,1} b_{-2}\ket{\Omega},
\end{equation}
and the gauge transformation of (\ref{tachyon_2}) up to level two is 
already given in \cite{WatiTalk} as
\begin{eqnarray}
\label{wati}
 \delta \eta_{0,1} & = & \mu_{2,1} (-\frac{16}{9} \eta_{0,1}-\frac{464}{243}
\eta_{2,1} +\frac{128}{81} \eta_{2,2}+\frac{1040}{243} \eta_{2,3})  
\nonumber \\
 \delta \eta_{2,1} & = & \mu_{2,1} (-3-\frac{176}{243} \eta_{0,1}
-\frac{11248}{6561}\eta_{2,1} -\frac{6016}{6561} \eta_{2,2}
+\frac{11440}{6561} \eta_{2,3})  
\nonumber \\ 
 \delta \eta_{2,2} & = & \mu_{2,1} (-1-\frac{224}{81} \eta_{0,1}
+\frac{992}{6561}\eta_{2,1} +\frac{1792}{729} \eta_{2,2}
+\frac{14560}{2187} \eta_{2,3})  
\nonumber \\ 
\label{level_trans}
 \delta \eta_{2,3} & = & \mu_{2,1} (1+\frac{80}{243}\eta_{0,1}
+\frac{2320}{6561} \eta_{2,1} -\frac{640}{2187} \eta_{2,2}
-\frac{9296}{6561} \eta_{2,3}),
\end{eqnarray}
which we have confirmed term by term.

On the other hand, we remind the reader that
$$
K_2 := L_2 - L_{-2} = L_2^m + L_2^g - L_{-2}^m - L_{-2}^g,
$$
where $L_m^g := \sum\limits_{n=-\infty}^{\infty} (2m-n) :b_n c_{m-n}: -
\delta_{m,0}$ is the ghost Virasoro operator with $:\;:$ being the
creation-annihilation normal ordering.
Recalling (\ref{tachyon_2}), we have
\begin{equation}
\label{K_2_level_tr}
K_2\ket{\Psi}=(3\eta_{2,1}+4\eta_{2,2}+13\eta_{2,3})\ket{\Omega}+
3\eta_{0,1}b_{-1} c_{-1} \ket{\Omega}+2\eta_{0,1}b_{-2} c_{0} \ket{\Omega}
-\eta_{0,1} L_{-2}^m\ket{\Omega}.
\end{equation}

We are now ready to check our proposal (\ref{unique_tachyon}) up to level 2
accuracy, i.e., can one tune the parameter $\mu_{2,1}$, so that
\begin{equation}
\label{check2}
K_2 \ket{\Psi_0} = \delta \ket{\Psi_0}
\end{equation}
would hold?

The left hand side of (\ref{check2}) is obtained by
substituting the numerical results of (\ref{level2_num}) into
(\ref{K_2_level_tr}):
$$
K_2 \ket{\Psi_0}=(0.11469, \; 1.1930, \; 0.79531, \; -0.39765).
$$
The right hand side of (\ref{check2}) is obtained via substitution of
(\ref{level2_num}) into (\ref{wati}):
$$
\delta \ket{\Psi_0}= \mu_{2,1}(-0.26656, \; -2.9785, \; -1.8485, \; 1.0238).
$$

Now we have 2 (Euclidean) vectors $(K_2 \ket{\Psi_0})_i$ and $(\delta
\ket{\Psi_0})_i$ of equal length which we wish to be
as close as possible if (\ref{unique_tachyon}) were to hold.
We subsequently choose the parameter $\mu_{2,1}$ by performing a
least-squares fit on these two vectors by minimizing the
Euclidean distance between the two.
$$
|K_2 \ket{\Psi_0}-\delta \ket{\Psi_0}| :=
\left( \sum\limits_i \left[ (K_2 \ket{\Psi_0})_i -  (\delta
\ket{\Psi_0})_i \right]^2 \right)^{\frac{1}{2}}.
$$

To this procedure we shall refer as ``best fit.'' At the present level
we arrive at
$$
\mu_{2,1}=-0.40732.
$$
Putting this value into $\delta \ket{\Psi_0}$ we get
$\delta \ket{\Psi_0}=(0.10857, \; 1.2132, \; 0.75290, \; -0.41702)$
and whence
$$
K_2 \ket{\Psi_0}-\delta \ket{\Psi_0}=(0.0061153, \; -0.020207,
\; 0.042406, \; 0.019368).
$$
A good estimator for our results is the normalized quantity, 
$$
\epsilon := \frac{|K_2 \ket{\Psi_0}-\delta \ket{\Psi_0}|}{|K_2
\ket{\Psi_0}|},
$$
which we wish to be as close to 0 as possible.
Using the above values, we have $\epsilon=0.034294$.
Therefore we conclude that up to level 2, our conjecture is 
accurate to $3.4\%$.
\subsection{Fitting at Level 4}
And thus we continue and to higher levels we shall go.
Now we keep the string field solution up to level
four and compare the two sides of  $K_2\ket{\Psi_0}$ and
$Q_{\Psi_0} \Lambda$ also up to level 4.

As we mentioned before, we can restrict the gauge parameters
to be of even levels as well, thus we can write $\Lambda$ as: 
\begin{equation}
\label{level4_gauge}
\begin{array}{rcl}
\ket{\Lambda} & = & \mu_{2,1} b_{-2}\ket{\Omega} +\mu_{4,1} b_{-4}\ket{\Omega}
  + \mu_{4,2}b_{-2} b_{-1} c_{-1} \ket{\Omega}  \\
& + & \mu_{4,3}b_{-3} b_{-1} c_0 \ket{\Omega}+\mu_{4,4}b_{-1} L_{-3}^m \ket{\Omega}
  + \mu_{4,5}b_{-2} L_{-2}^m \ket{\Omega},
\end{array}
\end{equation}
which has six numerical parameters.

Due to the overwhelming length of the gauge transformation and $K_2$
action on $\Psi_0$ to this level, we leave their presentation to
the Appendix. Again in accordance with our convention, we
can write the field into a vector with $14$ components in the order
$$
(\eta_{0,1},\eta_{2,i},\eta_{4,j}) \qquad (i=1,2,3; j=1,2,..,10).
$$
In this notation, the tachyon vacuum at level $(4,12)$ is given by
\begin{equation}
\label{num_level4}
\begin{array}{cc}
\ket{\Psi_0} = & (0.40072, \; -0.15029, \; 0, \; 0.041595, \; 0.041073, \;
	0.024192, \; 0.013691, \; \\
	& 0, \; 0, \; -0.0037419, \; 0, \; 0.0050132, \; 0, \; -0.00043064)\\
\end{array}
\end{equation}

We need now check (\ref{check2}) to level 4. 
The $K_2$ action on the left hand side is given by
\begin{equation}
\begin{array}{rcl}
K_2\ket{\Psi_0} & = & (0.089868, \; 1.2947, \; 0.75306, \; -0.42277, \; 
	0.75143, \; 0, \; -0.15029, \\
	& & 0, \; -0.30057, \; 0, \; 0, \; 0.27507, \; 0.083189, \; -0.041595)
\end{array}
\end{equation}
and $\delta \ket{\Psi_0}$ on the right hand side is a numerical 
function of the 6 $\mu$
parameters obtainable by substitution of (\ref{num_level4}) into the
appropriate expressions in the Appendix.

Again we minimize $|K_2\ket{\Psi_0}-\delta\ket{\Psi_0}|$ and find the
parameters as
\begin{eqnarray*}
\mu_{2, 1}= -0.54013,~~~&  \mu_{4, 1}= 0.18957,~~~ & \mu_{4, 2}= -0.37946, \\ 
  \mu_{4, 3} = -0.37645,~~~& \mu_{4, 4}= -0.12019,~~~  &\mu_{4, 5} = -0.022464
\end{eqnarray*}
Subsequently, we obtain
$$
\epsilon=\frac{|K_2 \ket{\Psi_0}-\delta \ket{\Psi_0}|}{|K_2 \ket{\Psi_0}|}
=0.016078.
$$
In conclusion then, the accuracy increases from $3.4\%$ at level 2 to 
$1.6\%$ at level 4.
\section{The Exactness of $K_1 \Psi_0$}
Having checked the validity of our conjecture (\ref{unique_tachyon})
for $K_2$ to within $1.6\%$, 
in this section we check if the $K_1$ action is a pure gauge
transformation. As we 
have mentioned in the beginning of the last section, we can restrict
the gauge parameters to odd levels only. 
Na\"{\i}vely the first nontrivial test is to expand $\ket{\Lambda}$ to
only level 1 which has 1 free parameter.
However, because to level 1 $K_1 \Psi_0$ has only 1
component, we would be lead to the trivial fitting of 1 parameter
to 1 constraint. Therefore we must start with level 3, by which we
mean that we expand $\Lambda$ to level 3 and $\Psi_0$ to level 2 and
thus $K_1 \Psi_0$ to level 3.
\subsection{Fitting at Level 3}
Up to level 3, we have four free parameters $\mu_{1,1}$ and $\mu_{3,i},
i=1,2,3$ in the gauge parameter:
$$
\ket{\Lambda}=\mu_{1,1} b_{-1}\ket{\Omega}+\mu_{3,1} b_{-3}\ket{\Omega}+
\mu_{3,2} b_{-2}b_{-1} c_0\ket{\Omega}+\mu_{3,3} b_{-1} L_{-2}^m\ket{\Omega}
$$
Once again the data of $\Psi_0$ to level 2 was given in 
(\ref{level2_num}).
The $K_1$ action and gauge transformation are subsequently presented
in the Appendix. Since $K_1 \Psi_0$ is at level 3, we have 6 fields in
the basis and a general field may be represented as
$(\eta_{1,1}, \eta_{3,i})$ with $(i=1,..,5)$.
Upon substitution of the numerical values in
(\ref{level2_num}), we have, to level 3,
$$
K_1\ket{\Psi_0}= (-0.25868,\; -0.41692,\; 0, \; 0, \; 0.040893, -0.040893).
$$

We perform the same procedure as in the previous section and minimize
$|K_1\ket{\Psi_0}-\delta\ket{\Psi_0}|$ to obtain the least-square
fitting parameters:
$$
\mu_{1, 1}= 0.88605,~~ \mu_{3,, 1}= -0.15821,~~\mu_{3,, 2}= 0.42491,~~ 
\mu_{3,, 3}= 0.23200.
$$
Consequently, the measure of our fit is given by
$$
\epsilon=\frac{|K_1 \ket{\Psi_0}-\delta \ket{\Psi_0}|}{|K_1 \ket{\Psi_0}|}
=0.036030
$$
Thus accuracy is achieved to within $3.6\%$, not so bad for this level.
\subsection{Fitting at Level 5}
To achieve greater accuracy, let us keep the string field up to level
5 and check (\ref{unique_tachyon}). Its two sides $K_1 \ket{\Psi_0}$
and $Q_{\Psi_0} \Lambda$ are both up to level 5, which in our notation
is a vector of length $22$, with 16 components at level 5 in addition
to those in the
previous subsection (indeed as remarked before, we need not include
the even levels):
$$
\left( \eta_{1,1}, \; \eta_{3,i}, \; \eta_{5,j} \right) 
\qquad (i=1,...,5; j=1,..16).
$$

In the same vein, we can restrict the gauge parameters
to odd levels only:
\begin{eqnarray*}
\label{level5_gauge}
\ket{\Lambda} & = & \mu_{1,1} b_{-1}\ket{\Omega} \\
& + & \mu_{3,1} b_{-3}\ket{\Omega} +
\mu_{3,2} b_{-2}b_{-1} c_0\ket{\Omega}+\mu_{3,3} b_{-1} L_{-2}^m\ket{\Omega}
\\
& + & \mu_{5,1} b_{-5} \ket{\Omega} + \mu_{5,2} b_{-2} b_{-1} c_{-2}
\ket{\Omega} +
\mu_{5,3} b_{-3} b_{-1} c_{-1} \ket{\Omega} \\
& + & \mu_{5,4} b_{-3} b_{-2} c_{0} \ket{\Omega} +
\mu_{5,5} b_{-4} b_{-1} c_{0} \ket{\Omega} +
\mu_{5,6} b_{-1} L^m_{-4} \ket{\Omega} \\
& + & \mu_{5,7} b_{-2} L^m_{-3} \ket{\Omega} +
\mu_{5,8} b_{-3} L^m_{-2} \ket{\Omega} +
\mu_{5,9} b_{-2} b_{-1} c_0 L^m_{-2} \ket{\Omega} \\
& + & \mu_{5,10} b_{-1} L^m_{-2} L^m_{-2} \ket{\Omega},
\end{eqnarray*}
which has 14 parameters $\mu$.

Once again, the gauge transformation and $K_1$ action on $\Psi_0$ can
be found in the Appendix. And thus equipped, the left hand side of 
(\ref{check2}) gives
$$
\begin{array}{rcl}
K_1 \ket{\Psi_0} & = & (-0.25043, \ -0.33721, \ 0.054765, \ -0.013691,
\ 0.021593, \\
& & -0.046608, \ 0.20537, \ 0.096767, \ 0.065265, \ 0.027382, \ 0, \\
& & 0.024192, \ 0.013691, \ -0.011656, \ 0.0037419, \ 0.0050132, \\
& & 0, \ 0.015040,  \ 0, \ 0, \ -0.00086128, \ 0.00043064).
\end{array}
$$

Finally we minimize $|K_1 \ket{\Psi_0} - \delta \ket{\Psi_0}|$ and
find the best-fit gauge parameters as:
$$
\begin{array}{lll}
\mu_{1, 1}= 0.96221, \ \ &  \mu_{3, 1}= -0.16665, \ \ & \mu_{3, 2}=
0.42762, \\
\mu_{3, 3} = 0.19259, \ \ & \mu_{5, 1}= -0.027057, \ \ &\mu_{5, 2} =
-1.2515, \\
\mu_{5, 3} = 0.31370, \ \ & \mu_{5, 4}= 1.0733, \ \ &\mu_{5, 5} =
-0.30612, \\
\mu_{5, 6} = -0.091788, \ \ & \mu_{5, 7}= 0.21383, \ \ &\mu_{5, 8} =
-0.30555, \\
\mu_{5, 9} = 0.19208, \ \ & \mu_{5, 10}= 0.050724,
\end{array}
$$
with an error estimate of:
$$
\epsilon=\frac{|K_1 \ket{\Psi_0} - \delta \ket{\Psi_0}|}{|K_1
\ket{\Psi_0}|}
=0.015128.
$$
So the accuracy increases from $3.6\%$ at level 3 to
$1.5\%$ at level 5.
\section{Concluding Remarks and Open questions}

Sen's second conjecture remains to be fully understood. A strong version of 
the conjecture states that the entire spectrum of the open string should
disappear from the perturbatively stable vacuum $\Psi_0$
and hence the cohomology
of $Q_{\Psi_0}$ should be trivial.
A reparameterization symmetry generated by $K_n$ in 
bosonic OSFT seems to be able to deform the tachyon vacuum whereby 
violating its uniqueness.
In this paper we have given a strong evidence in favor of the second
conjecture by explicitly showing that $K_n \Psi_0$ is merely a pure
gauge transformation and thus gives no new moduli to the tachyon vacuum.
Using a level truncation scheme, we have demonstrated that $K_{1,2}$ are
pure gauge up to level 5 (resp. level 4) to within an excellent accuracy of
$1.5\%$ (resp. $1.6\%$), and that all other $K_n$ are so by iteration.

Many open questions are of immediate interest for investigation; we list
a few here.
\begin{itemize}
\item An immediate check one could perform, as a test to the validity of
the level truncation procedure, is to see
to what accuracy is $K_n \Psi_0$ closed, 
i.e., though $Q_{\Psi_0} K_n \Psi_0$ should be identically zero,
level truncation spoils this and it would be interesting to check 
the numerics.

\item As we mentioned before, we can generate new solutions by acting 
$e^{\epsilon K_n}$ on a known solution. We can apply this method to, 
for example, lump solutions (\cite{0002117}-\cite{0101014})
and see what will happen. 
Indeed as is with $\Psi_0$, it is unlikely that $K_n$ will annihilate
the lump solution for all $n$, so we probably will obtain deformations 
of lumps.
The question is then to see if these new solutions are gauge equivalent to 
known lump solutions or if they do generate inequivalent new physical 
states.
If the answer is the latter, we would generate a part of the moduli space
to which the lumps belong. One particularly interesting example would be 
the solution generated by $e^{\epsilon K_1}$. Because $K_1$ changes the
level by one unit, by acting on the lumps we may obtain new solutions
which correspond to marginal deformations.

\item In this paper and in \cite{ellwood} only part of the cohomology of
$Q_{\Psi_0}$ is proven to be trivial. It will be very interesting to 
see if the entire cohomology is trivial. In other words,
if we have an
arbitrary deformation $\delta \Psi_0$ around the tachyon vacuum $\Psi_0$
which is closed $Q_{\Psi_0}\delta \Psi_0 = 0$, it must be exact, i.e.,
there exists a gauge parameter $\Lambda$ such that 
$Q_{\Psi_0} \Lambda = \delta \Psi_0$. 
One particular set of interesting deformations is those without 
momentum dependence because they are related to the possible moduli
space of translationally invariant solutions. When the solution is
unique, from a physical point of view, we should expect those
deformations to be in the trivial cohomology.
Proving the triviality of zero-momentum cohomology should
be readily tractable by level
truncation.

\item It is known that at the perturbative vacuum, $K_n$ is a good
symmetry of the theory. Indeed, $[K_n, Q_B] =0$. However, in the tachyon 
vacuum $\Psi_0$  we have 
$$
[K_n, Q_{\Psi_0}]A=(K_n\Psi_0) \star A-(-)^A A\star (K_n\Psi_0) \equiv
[K_n\Psi_0, A],
$$
which is not zero in general. This is in accord with \cite{0012251,0102112},
where the candidate BRST operators of the tachyon vacuum do not generally
commute with the $K_n$ operators\footnote{We thank B. Zwiebach for 
a discussion of this point.}. There may be a gauge  in which the tachyon vacuum
$\tilde{\Psi}_0$ satisfies
$K_n \tilde{\Psi}_0=0$ for all $n$, but we think this is unlikely. 
However, a subalgebra of $K_n$ might be a
symmetry of the tachyon vacuum. 
Any conclusions on these questions would have implications
for the SFT around the tachyon vacuum. 
\end{itemize}

\section*{Note added}
After the first version of this preprint was released, H. Hata sent us 
a formal proof that the $K_n \Psi_0$ are pure gauge. We thank him 
for pointing this out to us and, with his permission, we 
reproduce his proof here:
The proof uses the following three points:
\begin{enumerate}
\item[(1)] The $K_n$ can be expressed as an anticommutator:
$K_n = \{ Q_B, B_n \}$, with $B_n = b_n -(-1)^n b_{-n}$.
\item[(2)] The $B_n$ obey a Leibnitz rule for the star-product:
$B_n (A \star C) = (B_n A) \star C + (-1)^A A \star (B_n C)$.
\item[(3)] The equation of motion: $Q_B \Psi_0 + \Psi_0 \star \Psi_0 = 0$.
\end{enumerate}
Using the above, we can express $K_n \Psi_0$ in the following way:
\begin{eqnarray*}
K_n \Psi_0 & = & \{ Q_B, B_n \} \Psi_0 \\
& = & Q_B (B_n \Psi_0) + B_n (Q_B \Psi_0) \\
& = & Q_B (B_n \Psi_0) - B_n (\Psi_0 \star \Psi_0) \\ 
& = & Q_B (B_n \Psi_0) + \Psi_0 \star (B_n \Psi_0) - (B_n \Psi_0) 
\star \Psi_0 , 
\end{eqnarray*}
showing that (\ref{unique_tachyon}) holds, by taking $\Lambda = B_n \Psi_0$.

The work presented in this paper therefore reduces to a new check of the 
consistency of the level truncation method. The above proof also 
immediately answers our open question concerning deformations of 
lumps. Indeed, it can be seen from the proof, that for $K_n \Psi$ to 
be pure gauge, $\Psi$ only needs to be a solution of the equation of motion. 
The proof thus applies to a lump solution as well as to the vacuum.

Checking to what accuracy is $K_n \Psi_0$ closed, namely to see how 
well the property 
$Q_{\Psi_0}^2 = 0$ holds in the level truncation,  
would still be a good check 
of the level truncation. And of course, 
studying other parts of the cohomology, as well as looking for a subalgebra 
of the $K_n$ leaving the vacuum invariant, are still important open 
questions.

\acknowledgments
We would like to extend our sincere gratitude to B. Zwiebach for his
many insightful comments as well as careful proof-reading and
corrections of the
manuscript. Furthermore we would like to thank
I. Ellwood, N. Prezas, L. Rastelli, A. Sen, J. S. Song and W. Taylor for 
valuable discussions.
And we are indebted to H.Hata for sharing with us the proof presented in the 
note added.

\vspace{4.5cm}

\appendix
\section{Appendix}
In this Appendix we shall tabulate the details used in our
calculations. In subsections A.1 and A.2 we present the basis of the
fields for ghost numbers 0 and 1, In A.3, we present the action of
$K_1$ and $K_2$ on the string field theory vacuum to level 4. Finally
in subsections A.4 and A.5 we present the gauge transformations of the
vacuum to level 5.
~\\
~\\
~\\
\subsection{The Basis of Ghost Number 1 Fields}
As $\Psi_0$ is ghost number 1, we here tabulate the basis of the
ghost number 1 fields up to level 5, consisting of a total of 14 in even
levels and 22 in odd levels.
The numerical parameters $\eta_{\ell , i}$ denote the expansion coefficient
of the field  $\Psi$ at the $i$-th field at level $\ell$. For the
vacuum these parameters have been computed to great precision in
\cite{0002237}; we use their results at level $(4,12)$.
~\\
~\\
\par
\begin{tabular}{|l|l|l|l|}  \hline
Level   &   Field    &  Coefficient   & vev at level (4,12) \\  \hline
 0  & $\ket{\Omega}=c_1\ket{0}$  & $\eta_{0,1}$  & 0.40072 
\\  \hline \hline

 1  & $b_{-1} c_0\ket{\Omega}$ & $\eta_{1,1}$ & 0
\\  \hline \hline

 2 \ (3 fields)  & $b_{-1} c_{-1} \ket{\Omega}$ & $\eta_{2,1}$ &  -0.15029 
\\  \hline
	&  $b_{-2} c_{0} \ket{\Omega}$ & $\eta_{2,2}$ & 0\\  \hline
	& $L_{-2}^m\ket{\Omega}$  & $\eta_{2,3}$ & 0.041595  \\
\hline \hline

 3 \ (5 fields)  & $b_{-1} c_{-2} \ket{\Omega}$ & $\eta_{3,1}$ & 0 \\  \hline
	& $b_{-2} c_{-1} \ket{\Omega}$ & $\eta_{3,2}$ & 0 \\  \hline
	& $b_{-3} c_{0} \ket{\Omega}$ & $\eta_{3,3}$ & 0 \\  \hline
	& $L_{-3}^m\ket{\Omega}$  & $\eta_{3,4}$ &  0 \\  \hline
	& $b_{-1} c_0 L_{-2}^m \ket{\Omega}$  & $\eta_{3,5}$ & 0 \\
\hline \hline

 4 \ (10 fields)  & $b_{-1} c_{-3} \ket{\Omega}$ & $\eta_{4,1}$
	&0.041073 \\  \hline
	& $b_{-2} c_{-2} \ket{\Omega}$ & $\eta_{4,2}$ &  0.024192  \\  \hline 
	& $b_{-3} c_{-1} \ket{\Omega}$ & $\eta_{4,3}$ & 0.013691  \\  \hline 
	& $b_{-4} c_{0} \ket{\Omega}$ & $\eta_{4,4}$ & 0 \\  \hline 
	& $b_{-2} b_{-1} c_{-1} c_0\ket{\Omega}$ &  $\eta_{4,5}$ & 0\\  \hline 
	& $L_{-4}^m\ket{\Omega}$ &  $\eta_{4,6}$ & -0.0037419\\  \hline 
	& $b_{-1} c_0 L_{-3}^m \ket{\Omega}$ &  $\eta_{4,7}$ & 0 \\  \hline 
	& $b_{-1} c_{-1} L_{-2}^m  \ket{\Omega}$ &  $\eta_{4,8}$ & 
		0.0050132 \\  \hline 
	& $b_{-2} c_{0} L_{-2}^m  \ket{\Omega}$ &  $\eta_{4,9}$ & 0 \\  \hline 
	& $L_{-2}^m L_{-2}^m  \ket{\Omega}$ &  $\eta_{4,10}$ &
 		-0.00043064  \\  \hline \hline
\end{tabular}

\begin{tabular}{|l|l|l|l|}  \hline
Level   &   Field    &  Coefficient   & vev at level (4,12) \\  \hline
5 \ (16 fields)  & $b_{-1} c_{-4} \ket{\Omega}$ & $\eta_{5,1}$ & 0 \\  \hline
	& $b_{-2} c_{-3} \ket{\Omega}$ & $\eta_{5,2}$ & 0\\  \hline
	& $b_{-3} c_{-2} \ket{\Omega}$ & $\eta_{5,3}$ & 0 \\  \hline
	& $b_{-4} c_{-1} \ket{\Omega}$ & $\eta_{5,4}$ & 0    \\   \hline
	& $b_{-5} c_{0} \ket{\Omega}$ & $\eta_{5,5}$ & 0  \\   \hline
	& $b_{-2} b_{-1} c_{-2} c_0\ket{\Omega}$ &  $\eta_{5,6}$ & 0  
  		\\   \hline
	& $b_{-3} b_{-1} c_{-1} c_0\ket{\Omega}$ &  $\eta_{5,7}$ & 0  
  		\\   \hline
	& $L_{-5}^m \ket{\Omega}$ &  $\eta_{5,8}$ & 0   \\  \hline 
	& $b_{-1} c_0 L_{-4}^m \ket{\Omega}$ &  $\eta_{5,9}$ & 0 
		\\  \hline 
	&  $b_{-1} c_{-1} L_{-3}^m \ket{\Omega}$ & $\eta_{5,10}$ &  0
		\\  \hline 
	& $b_{-2} c_{0}  L_{-3}^m \ket{\Omega}$ & $\eta_{5,11}$ & 0 
		\\  \hline 
	& $b_{-1} c_{-2}  L_{-2}^m \ket{\Omega}$ & $\eta_{5,12}$ & 0
		\\  \hline 
	& $b_{-2} c_{-1} L_{-2}^m\ket{\Omega}$ & $\eta_{5,13}$ & 0 
		\\  \hline
	& $b_{-3} c_{0}  L_{-2}^m\ket{\Omega}$ & $\eta_{5,14}$ & 0 
		\\   \hline
	& $L_{-3}^m L_{-2}^m \ket{\Omega}$ & $\eta_{5,15}$ & 0 
		\\  \hline
	& $b_{-1} c_0 L_{-2}^m L_{-2}^m  \ket{\Omega}$ &  
		$\eta_{5,16}$ & 0 \\  \hline 
\end{tabular}

\subsection{The Basis of Ghost Number 0 Fields}
The gauge transformation parameter $\ket{\Lambda}$ is of ghost number
0, thus we here present the basis for ghost number 0 fields. Analogous
to the previous subsection, we use $\mu_{\ell , i}$ for $\ell = 1,..,5$,
and $i$ indexing within each level
to denote the coefficient of the expansion of $\ket{\Lambda}$ into the
basis. A least-squares fit was then performed in order to minimize
the difference between the $K$ action on the vacuum and the gauge
transformation therefrom. Below, the columns Fit $n$ refer to the
solution of the parameters $\mu$ at the best-fit at level $n$.
\paragraph{}
\begin{tabular}{|l|l|l|l|l|l|l|} 
\hline
Level  &  Field  & Coefficient  & Fit 2 & Fit 3 & Fit 4 & Fit 5 
	\\  \hline  
1	& $b_{-1}\ket{\Omega}$ &  $\mu_{1,1}$ & & 0.886 & & 0.962
	\\ \hline \hline
2 	& $b_{-2}\ket{\Omega}$ &$\mu_{2,1}$ & -0.407 & & -0.540 &
	\\ \hline \hline
3	& $b_{-3}\ket{\Omega}$ &$\mu_{3,1}$ &  & -0.158 & & -0.167
		\\ \hline
	& $b_{-2} b_{-1} c_0 \ket{\Omega}$ &$\mu_{3,2}$ & & 0.425 & & 0.428
		\\ \hline
	& $b_{-1} L_{-2}^m \ket{\Omega}$ &$\mu_{3,3}$ & & 0.232 & & 0.193
		\\ \hline \hline
4	& $b_{-4}\ket{\Omega}$ &$\mu_{4,1}$ &  & &0.190 &
		\\ \hline
	& $b_{-2} b_{-1} c_{-1} \ket{\Omega}$ &$\mu_{4,2}$ & & &-0.379 &
		\\ \hline
	& $b_{-3} b_{-1} c_0 \ket{\Omega}$ &$\mu_{4,3}$ & & &-0.376 &
		\\ \hline
	& $b_{-1} L_{-3}^m \ket{\Omega}$ &$\mu_{4,4}$ & & &-0.120 &
		\\ \hline
	& $b_{-2} L_{-2}^m \ket{\Omega}$ &$\mu_{4,5}$ & & &-0.0225 &
		\\ \hline \hline
5	& $b_{-5}\ket{\Omega}$ &$\mu_{5,1}$ &  & & & -0.0271
		\\ \hline
	& $b_{-2} b_{-1} c_{-2} \ket{\Omega}$ &$\mu_{5,2}$ & & & & -1.25
		\\ \hline
	& $b_{-3} b_{-1} c_{-1} \ket{\Omega}$ &$\mu_{5,3}$ & & & & 0.314
		\\ \hline
	& $b_{-3} b_{-2} c_0 \ket{\Omega}$ &$\mu_{5,4}$ & & & & 1.07
		\\ \hline
	& $b_{-4} b_{-1} c_0 \ket{\Omega}$ &$\mu_{5,5}$ & & & & -0.306
		\\ \hline
	& $b_{-1} L_{-4}^m \ket{\Omega}$ &$\mu_{5,6}$ & & & & -0.0918
		\\ \hline
	& $b_{-2} L_{-3}^m \ket{\Omega}$ &$\mu_{5,7}$ & & & & 0.214
		\\ \hline
	& $b_{-3} L_{-2}^m \ket{\Omega}$ &$\mu_{5,8}$ & & & & -0.306
		\\ \hline
	& $b_{-2} b_{-1} c_{0} L_{-2}^m  \ket{\Omega}$ &$\mu_{5,9}$ &
		& & & 0.192 \\ \hline
	& $ b_{-1} L_{-2}^m L_{-2}^m  \ket{\Omega}$ &$\mu_{5,10}$ & &
		& & 0.0507
		\\ \hline
\end{tabular}
\subsection{$K_1$ and $K_2$ Actions on $\ket{\Psi_0}$}
We act  $K_1$ and $K_2$ on the vacuum $\Psi_0$
(only the action on nonzero components of $\Psi_0$ is kept):
\begin{eqnarray*}
K_1 \Psi_0 & = & [(-\eta_{0,1}-\eta_{2,1})b_{-1} c_0 \ket{\Omega} ]
 +  [(3\eta_{2,1}+\eta_{4,1}+3\eta_{4,2}) b_{-1} c_{-2} \ket{\Omega} \\  
& + & (4\eta_{4,3}) b_{-2} c_{-1} \ket{\Omega}
      + (-\eta_{4,3}) b_{-3} c_0  \ket{\Omega}  \\
& + & (\eta_{2,3}+5\eta_{4,6}+3 \eta_{4,10}) L_{-3}^m \ket{\Omega}
     +(-\eta_{2,3}-\eta_{4,8})b_{-1} c_0 L_{-2}^m  \ket{\Omega}] \\
& + &[ (5\eta_{4,1}) b_{-1} c_{-4}\ket{\Omega}+ (4\eta_{4,2}) b_{-2} c_{-3} \ket{\Omega}
     +(\eta_{4,2}+3\eta_{4,3}) b_{-3} c_{-2}\ket{\Omega}  \\
& + & (2\eta_{4,3}) b_{-4} c_{-1}\ket{\Omega} +(\eta_{4,2}) b_{-2} b_{-1} c_{-2}
      c_0 \ket{\Omega}+ (\eta_{4,3}) b_{-3} b_{-1} c_{-1}  c_0 \ket{\Omega} \\
& + & (3\eta_{4,6}+\eta_{4,10}) L_{-5}^m \ket{\Omega} -(\eta_{4,6}) b_{-1} c_0
      L_{-4}^m \ket{\Omega} +(\eta_{4,8}) b_{-1} c_{-1} L_{-3}^m \ket{\Omega}  \\
& + & (3\eta_{4,8})b_{-1} c_{-2} L_{-2}^m \ket{\Omega} + (2\eta_{4,10})
      L_{-3}^m L_{-2}^m  \ket{\Omega}-(\eta_{4,10}) b_{-1} c_0 L_{-2}^m L_{-2}^m
      \ket{\Omega}]
\end{eqnarray*}

\par

\begin{eqnarray*}
K_2 \Psi_0 & = & [(3\eta_{2,1}+13\eta_{2,3})\ket{\Omega} ] \\
& + & [(3\eta_{0,1}-\eta_{4,1}+5\eta_{4,3}+13\eta_{4,8}) b_{-1} c_{-1} \ket{\Omega} 
      +(2\eta_{0,1}-2\eta_{4,2}) b_{-2} c_0 \ket{\Omega}  \\
& + & (-\eta_{0,1}+6\eta_{4,6}+3\eta_{4,8}+34\eta_{4,10}) L_{-2}^m \ket{\Omega} ]
      + [(-5\eta_{2,1}) b_{-1} c_{-3} \ket{\Omega} \\
& + & \eta_{2,1} b_{-3} c_{-1} \ket{\Omega}+(2\eta_{2,1}) b_{-2} b_{-1}
      c_{-1} c_0 \ket{\Omega}+(-\eta_{2,1} +3\eta_{2,3}) b_{-1} c_{-1} L_{-2}^m
      \ket{\Omega} \\
& + & (2\eta_{2,3}) b_{-2} c_0 L_{-2}^m \ket{\Omega} -\eta_{2,3}
      L_{-2}^m L_{-2}^m \ket{\Omega}]
\end{eqnarray*}
\subsection{Gauge Transformation of the Even Level String Field}
Let us present the heuristics of the computation required in the gauge
transformation $\delta \Psi := Q_B \Lambda + \Psi \star \Lambda -
\Lambda \star \Psi$. The only non-trivial part is the computation of the
$\star$-product.
Since we are working under a level-truncation scheme, to compute $B
\star C$ for string fields $B$ and $C$, it suffices to
find, level-by-level, the coefficients of the expansion of the
star-product into the basis of each level, i.e.,
$$
B \star C = \sum\limits_{\ell,i} x_{\ell,i} \psi_{\ell,i},
$$
with $\psi_{\ell,i}$ the $i$-th field basis at level $\ell$ and
$x_{\ell,i}$ the coefficients we wish to determine.
Defining the orthonormal basis
 $\tilde{\psi}_{\ell,i}$, so that
$$
\langle \tilde{\psi}_{\ell,i}, \psi_{\ell',i'} \rangle = \delta_{\ell
\ell'} \delta_{i i'},
$$
where $\langle \cdot ,\cdot\rangle$ is the BPZ inner product, we arrive at
$x_{\ell,i} = \langle \tilde{\psi}_{\ell,i}, B \star C \rangle,$
which simplifies by the definition
$\langle A, B \star C \rangle := \langle A, B, C \rangle,$
to
$$
x_{\ell,i} = \langle \tilde{\psi}_{\ell,i}, B, C \rangle.
$$

For an example, let us determine the coefficient $x$ in
$$
\ket{\Omega} \star b_{-2}\ket{\Omega} = x \ket{\Omega} + \ldots
$$
The orthogonal state to $\ket{\Omega}$ is $c_0 \ket{\Omega}$,
therefore $x = \langle c_0 \ket{\Omega}, \ket{\Omega},
b_{-2} \ket{\Omega} \rangle = -\frac{8}{9}$ in a normalization where 
$\langle \ket{\Omega}, \ket{\Omega}, \ket{\Omega} \rangle = 3$ in
accordance with \cite{0002237}. The computation of the 3-correlator we
leave the reader to a vast literature \cite{9912249,0002237,LPP,0006240}. 
As another example, let us compute
$$
b_{-1}c_{-1} \ket{\Omega} \star b_{-2}\ket{\Omega} = x b_{-2}c_0
	\ket{\Omega} + \ldots.
$$
The orthogonal state to $b_{-2}c_0\ket{\Omega}$ is $c_{-2}\ket{\Omega}$,
whence $x = \langle c_{-2} \ket{\Omega}, b_{-1}c_{-1} \ket{\Omega}, 
b_{-2}\ket{\Omega} \rangle = \frac{496}{6561}$.

We point out further that
a simplification is at hand due to the relation:
\begin{equation}
\label{simplify}
\langle A, B, C \rangle = (-1)^{1 + g(A)g(B) + \ell (A) + \ell (B) +
\ell (C)} \langle A, C, B \rangle,
\end{equation}
where $g(X)$ and $\ell (X)$ are the ghost number and level of $X$
respectively (we take $g(\ket{\Omega})=1$).

This simplification (\ref{simplify}) is crucial to the observations in
the second paragraph at the beginning of Section 3.
We need to compute $\Phi\star \Lambda-\Lambda \star \Phi$, so we expand
it into the basis $A$ and the coefficients are 
\[
\langle A,\Phi,\Lambda \rangle - \langle A,\Lambda, \Phi \rangle
= \langle A,\Phi,\Lambda \rangle (1+(-)^{g(A)g(\Phi)+\ell (A)+\ell
(\Phi)+\ell (\Lambda)})
\]
In our case, we have always that $g(A)=2$, so we must have 
$$
\ell (A)+\ell (\Phi)+\ell (\Lambda)=\mbox{even};
$$
otherwise the coefficient would be zero.

For example, when $\ket{\Phi}=\ket{\Omega}$ and
$\ket{\Lambda}=b_{-2}\ket{\Omega}$,
only even levels of $A$ have non zero coefficients, while when 
$\ket{\Phi}=\ket{\Omega}$ and $\ket{\Lambda}=b_{-1}\ket{\Omega}$, only
the odd levels of $A$ have non zero coefficients. 
Of such a simplification we have taken great 
advantage in the computations of Sections 3 and 4.

We present below the gauge transformation on a string
field. Here we consider the case that the string field
has only even levels, so for the gauge transformation of
even levels we have only even level gauge parameters while
for the gauge transformation of odd levels we have only
odd level gauge parameters. 
We divide the gauge transformation into two parts. The first part 
($\delta^{(1)} \eta_{\ell, i}$) is 
$Q_B \Lambda$, which is exact at every level.
The second part ($\delta \eta_{\ell, i}$) 
is $\Psi_0 \star \Lambda-\Lambda \star \Psi_0$; it is 
approximate in the level truncation.

\vspace{0.5cm}

\par
{\bf $Q_B \Lambda$ part:}
\begin{eqnarray*}
\delta^{(1)} \eta_{2,1}  & = & -3 \mu_{2,1}  \\[-0.2cm] 
\delta^{(1)} \eta_{2,2}  & = & - \mu_{2,1}  \\[-0.2cm]
\delta^{(1)} \eta_{2,3}  & = & 1 \mu_{2,1}  \\[-0.2cm]
\delta^{(1)} \eta_{4,1}  & = & -7\mu_{4,1} +5\mu_{4,2}
+6\mu_{4,3}-52\mu_{4,4} \\[-0.2cm]
\delta^{(1)} \eta_{4,2}  & = & -6 \mu_{4,1}-3\mu_{4,2}-13 \mu_{4,5}  \\[-0.2cm]
\delta^{(1)} \eta_{4,3}  & = & -5 \mu_{4,1}-1\mu_{4,2}-2\mu_{4,3} \\[-0.2cm]
\delta^{(1)} \eta_{4,4}  & = & -3 \mu_{4,1}-2\mu_{4,3} \\[-0.2cm]
\delta^{(1)} \eta_{4,5}  & = & -3\mu_{4,2}+4\mu_{4,3}  \\[-0.2cm]
\delta^{(1)} \eta_{4,6}  & = &  \mu_{4,1}+2\mu_{4,4} \\[-0.2cm]
\delta^{(1)} \eta_{4,7}  & = & \mu_{4,3}-3\mu_{4,4} \\[-0.2cm]
\delta^{(1)} \eta_{4,8}  & = & \mu_{4,2}-4\mu_{4,4}-3\mu_{4,5}\\[-0.2cm]
\delta^{(1)} \eta_{4,9}  & = & -3\mu_{4,5}\\[-0.2cm]
\delta^{(1)} \eta_{4,10}  & = &  \mu_{4,5} \\[-0.2cm]
\end{eqnarray*}
for even level and 
\begin{eqnarray*}
\delta^{(1)} \eta_{3,1}  & = & -5\mu_{3,1}+4\mu_{3,2}-13\mu_{3,3} \\[-0.2cm]
\delta^{(1)} \eta_{3,2}  & = & -4\mu_{3,1}-2\mu_{3,2} \\[-0.2cm]
\delta^{(1)} \eta_{3,3}  & = & -2\mu_{3,1}-\mu_{3,2} \\[-0.2cm]
\delta^{(1)} \eta_{3,4}  & = & \mu_{3,1}+\mu_{3,3} \\[-0.2cm]
\delta^{(1)} \eta_{3,5}  & = & \mu_{3,2}-2\mu_{3,3} \\[-0.2cm]
\delta^{(1)} \eta_{4,1}  & = &
-9\mu_{5,1}+6\mu_{5,2}+7\mu_{5,3}+8\mu_{5,5}
     -130\mu_{5,6}-78 \mu_{5,10}  \\[-0.2cm]
\delta^{(1)} \eta_{4,2}  & = &
-8\mu_{5,1}-4\mu_{5,2}+6\mu_{5,4}-52\mu_{5,6} \\[-0.2cm]
\delta^{(1)} \eta_{4,3}  & = &-7\mu_{5,1}-\mu_{5,2}-3\mu_{5,3}-4\mu_{5,4}
    -13\mu_{5,8} \\[-0.2cm]
\delta^{(1)} \eta_{4,4}  & = &-6\mu_{5,1}-2\mu_{5,3}-2\mu_{5,5} \\[-0.2cm]
\delta^{(1)} \eta_{4,5}  & = &-4\mu_{5,1}-\mu_{5,4}-3\mu_{5,5} \\[-0.2cm]
\delta^{(1)} \eta_{4,6}  & =
&-4\mu_{5,2}-5\mu_{5,4}+6\mu_{5,5}+13\mu_{5,9} \\[-0.2cm]
\delta^{(1)} \eta_{4,7}  & = &-4\mu_{5,3}+3\mu_{5,4}+5\mu_{5,5}  \\[-0.2cm]
\delta^{(1)} \eta_{4,8}  & = &\mu_{5,1}+3\mu_{5,6}+\mu_{5,7}+\mu_{5,10}
\\[-0.2cm]
\delta^{(1)} \eta_{4,9}  & = &\mu_{5,5}-4\mu_{5,6}  \\[-0.2cm]
\delta^{(1)} \eta_{4,10}  & = &\mu_{5,3}-5\mu_{5,6}-3\mu_{5,7}-3\mu_{5,10}
\\[-0.2cm]
\delta^{(1)} \eta_{4,11}  & = &\mu_{5,4}-4\mu_{5,7}-\mu_{5,9}  \\[-0.2cm]
\delta^{(1)} \eta_{4,12}  & = &\mu_{5,2}-6\mu_{5,6}-5\mu_{5,8}+4\mu_{5,9}
   -34\mu_{5,10}  \\[-0.2cm]
\delta^{(1)} \eta_{4,13}  & = & -4\mu_{5,7}-4\mu_{5,8}-2\mu_{5,9}  \\[-0.2cm]
\delta^{(1)} \eta_{4,14}  & = & -\mu_{5,4}-4\mu_{5,8}-\mu_{5,9} \\[-0.2cm]
\delta^{(1)} \eta_{4,15}  & = &\mu_{5,7}+\mu_{5,8}+2\mu_{5,10}  \\[-0.2cm]
\delta^{(1)} \eta_{4,16}  & = &\mu_{5,9}-4\mu_{5,10}
\end{eqnarray*}
for odd level (only nonzero contributions are listed).

\vspace{0.5cm}

\par{\bf $\Psi_0\star \Lambda-\Lambda \star \Psi_0$ part:}

\vspace{0.2cm}

Here we show only $\delta \eta_{0,1}$ and $\delta \eta_{1,1}$. 
For the complete results up to levels 4 and 5 for all $\eta$'s,
due to the enormity of the expressions,
the reader is referred to the web-page \\
\href{http://pierre.mit.edu/~yhe/gaugetransf.dvi}{http://pierre.mit.edu/~yhe/gaugetransf.dvi}.
\begin{eqnarray*}
\delta \eta_{0,1} & = &\frac{-16 \eta_{0,1} \mu_{2,1}}{9} - 
  \frac{464 \eta_{2,1} \mu_{2,1}}{243} + 
  \frac{128 \eta_{2,2} \mu_{2,1}}{81} + 
  \frac{1040 \eta_{2,3} \mu_{2,1}}{243} 
\\& & - 
  \frac{8576 \eta_{4,1} \mu_{2,1}}{6561} + 
  \frac{496 \eta_{4,2} \mu_{2,1}}{729} + 
  \frac{7040 \eta_{4,3} \mu_{2,1}}{6561} - 
  \frac{2816 \eta_{4,4} \mu_{2,1}}{2187}
\\& & + 
  \frac{6016 \eta_{4,5} \mu_{2,1}}{6561} - 
  \frac{2080 \eta_{4,6} \mu_{2,1}}{243} + 
  \frac{30160 \eta_{4,8} \mu_{2,1}}{6561} - 
  \frac{8320 \eta_{4,9} \mu_{2,1}}{2187} 
\\& &- 
  \frac{112736 \eta_{4,10} \mu_{2,1}}{6561} + 
  \frac{352 \eta_{0,1} \mu_{4,1}}{243} + 
  \frac{6112 \eta_{2,1} \mu_{4,1}}{6561} - 
  \frac{2816 \eta_{2,2} \mu_{4,1}}{2187} 
\\& &- 
  \frac{22880 \eta_{2,3} \mu_{4,1}}{6561} - 
  \frac{290560 \eta_{4,1} \mu_{4,1}}{177147} + 
  \frac{32864 \eta_{4,2} \mu_{4,1}}{177147} - 
  \frac{9472 \eta_{4,3} \mu_{4,1}}{19683} 
\\& &+ 
  \frac{61952 \eta_{4,4} \mu_{4,1}}{59049} - 
  \frac{7424 \eta_{4,5} \mu_{4,1}}{19683} + 
  \frac{45760 \eta_{4,6} \mu_{4,1}}{6561} - 
  \frac{397280 \eta_{4,8} \mu_{4,1}}{177147}
\\& & + 
  \frac{183040 \eta_{4,9} \mu_{4,1}}{59049} + 
  \frac{2480192 \eta_{4,10} \mu_{4,1}}{177147} + 
  \frac{176 \eta_{0,1} \mu_{4,2}}{243} + 
  \frac{11248 \eta_{2,1} \mu_{4,2}}{6561} 
\\& &+ 
  \frac{6016 \eta_{2,2} \mu_{4,2}}{6561} - 
  \frac{11440 \eta_{2,3} \mu_{4,2}}{6561} + 
  \frac{17536 \eta_{4,1} \mu_{4,2}}{19683} + 
  \frac{217136 \eta_{4,2} \mu_{4,2}}{177147}
\\& & + 
  \frac{14720 \eta_{4,3} \mu_{4,2}}{177147} - 
  \frac{7424 \eta_{4,4} \mu_{4,2}}{19683} - 
  \frac{80512 \eta_{4,5} \mu_{4,2}}{177147} + 
  \frac{22880 \eta_{4,6} \mu_{4,2}}{6561}
\\& & - 
  \frac{731120 \eta_{4,8} \mu_{4,2}}{177147} - 
  \frac{391040 \eta_{4,9} \mu_{4,2}}{177147} + 
  \frac{1240096 \eta_{4,10} \mu_{4,2}}{177147} + 
  \frac{8192 \eta_{2,1} \mu_{4,3}}{6561}
\\& & + 
  \frac{303104 \eta_{4,1} \mu_{4,3}}{177147} + 
  \frac{131072 \eta_{4,2} \mu_{4,3}}{177147} - 
  \frac{139264 \eta_{4,3} \mu_{4,3}}{177147} - 
  \frac{532480 \eta_{4,8} \mu_{4,3}}{177147} 
\\& &+ 
  \frac{212992 \eta_{2,3} \mu_{4,4}}{6561} + 
  \frac{2129920 \eta_{4,6} \mu_{4,4}}{19683} - 
  \frac{13631488 \eta_{4,7} \mu_{4,4}}{177147} - 
  \frac{4046848 \eta_{4,8} \mu_{4,4}}{177147}
\\& & - 
  \frac{1703936 \eta_{4,9} \mu_{4,4}}{177147} - 
  \frac{20873216 \eta_{4,10} \mu_{4,4}}{177147} + 
  \frac{1040 \eta_{0,1} \mu_{4,5}}{243} + 
  \frac{30160 \eta_{2,1} \mu_{4,5}}{6561} 
\\& &- 
  \frac{8320 \eta_{2,2} \mu_{4,5}}{2187} - 
  \frac{120848 \eta_{2,3} \mu_{4,5}}{6561} + 
  \frac{557440 \eta_{4,1} \mu_{4,5}}{177147} - 
  \frac{32240 \eta_{4,2} \mu_{4,5}}{19683}
\\& & - 
  \frac{457600 \eta_{4,3} \mu_{4,5}}{177147} + 
  \frac{183040 \eta_{4,4} \mu_{4,5}}{59049} - 
  \frac{391040 \eta_{4,5} \mu_{4,5}}{177147} + 
  \frac{3117920 \eta_{4,6} \mu_{4,5}}{177147}
\\& & - 
  \frac{1703936 \eta_{4,7} \mu_{4,5}}{177147} - 
  \frac{3504592 \eta_{4,8} \mu_{4,5}}{177147} + 
  \frac{966784 \eta_{4,9} \mu_{4,5}}{59049} + 
  \frac{5034016 \eta_{4,10} \mu_{4,5}}{59049}
\end{eqnarray*}

\begin{eqnarray*}
\delta \eta_{1,1} & = & 
 \frac{-16 \eta_{0,1} \mu_{1,1}}{9} + 
  \frac{16 \eta_{2,1} \mu_{1,1}}{81} + 
  \frac{896 \eta_{2,2} \mu_{1,1}}{243} + 
  \frac{1040 \eta_{2,3} \mu_{1,1}}{243} \\ & &  - 
  \frac{640 \eta_{4,1} \mu_{1,1}}{2187} + 
  \frac{5488 \eta_{4,2} \mu_{1,1}}{6561} - 
  \frac{640 \eta_{4,3} \mu_{1,1}}{729} - 
  \frac{15616 \eta_{4,4} \mu_{1,1}}{6561} \\ & & - 
  \frac{7808 \eta_{4,5} \mu_{1,1}}{6561} - 
  \frac{2080 \eta_{4,6} \mu_{1,1}}{243} - 
  \frac{1040 \eta_{4,8} \mu_{1,1}}{2187} - 
  \frac{58240 \eta_{4,9} \mu_{1,1}}{6561} \\ & & - 
  \frac{112736 \eta_{4,10} \mu_{1,1}}{6561} - 
  \frac{80 \eta_{0,1} \mu_{3,1}}{81} + 
  \frac{80 \eta_{2,1} \mu_{3,1}}{729} - 
  \frac{7040 \eta_{2,2} \mu_{3,1}}{6561} \\ & & + 
  \frac{5200 \eta_{2,3} \mu_{3,1}}{2187} - 
  \frac{159872 \eta_{4,1} \mu_{3,1}}{177147} + 
  \frac{7600 \eta_{4,2} \mu_{3,1}}{6561} - 
  \frac{3200 \eta_{4,3} \mu_{3,1}}{6561} \\ & & + 
  \frac{9472 \eta_{4,4} \mu_{3,1}}{19683} + 
  \frac{108160 \eta_{4,5} \mu_{3,1}}{177147} - 
  \frac{10400 \eta_{4,6} \mu_{3,1}}{2187} - 
  \frac{5200 \eta_{4,8} \mu_{3,1}}{19683} \\ & & + 
  \frac{457600 \eta_{4,9} \mu_{3,1}}{177147} - 
  \frac{563680 \eta_{4,10} \mu_{3,1}}{59049} - 
  \frac{256 \eta_{0,1} \mu_{3,2}}{243} - 
  \frac{9472 \eta_{2,1} \mu_{3,2}}{6561} \\ & & + 
  \frac{8192 \eta_{2,2} \mu_{3,2}}{6561} + 
  \frac{16640 \eta_{2,3} \mu_{3,2}}{6561} - 
  \frac{114688 \eta_{4,1} \mu_{3,2}}{177147} - 
  \frac{6400 \eta_{4,2} \mu_{3,2}}{177147}  \\ & &+ 
  \frac{81920 \eta_{4,3} \mu_{3,2}}{177147} - 
  \frac{114688 \eta_{4,4} \mu_{3,2}}{177147} + 
  \frac{303104 \eta_{4,5} \mu_{3,2}}{177147} - 
  \frac{33280 \eta_{4,6} \mu_{3,2}}{6561} \\ & & + 
  \frac{615680 \eta_{4,8} \mu_{3,2}}{177147} - 
  \frac{532480 \eta_{4,9} \mu_{3,2}}{177147} - 
  \frac{1803776 \eta_{4,10} \mu_{3,2}}{177147} + 
  \frac{1040 \eta_{0,1} \mu_{3,3}}{243}  \\ & &- 
  \frac{1040 \eta_{2,1} \mu_{3,3}}{2187} - 
  \frac{58240 \eta_{2,2} \mu_{3,3}}{6561} - 
  \frac{120848 \eta_{2,3} \mu_{3,3}}{6561} + 
  \frac{41600 \eta_{4,1} \mu_{3,3}}{59049} \\ & & - 
  \frac{356720 \eta_{4,2} \mu_{3,3}}{177147} + 
  \frac{41600 \eta_{4,3} \mu_{3,3}}{19683} + 
  \frac{1015040 \eta_{4,4} \mu_{3,3}}{177147} + 
  \frac{507520 \eta_{4,5} \mu_{3,3}}{177147} \\ & & + 
  \frac{3117920 \eta_{4,6} \mu_{3,3}}{177147} - 
  \frac{1703936 \eta_{4,7} \mu_{3,3}}{177147} + 
  \frac{120848 \eta_{4,8} \mu_{3,3}}{59049}  + 
  \frac{6767488 \eta_{4,9} \mu_{3,3}}{177147} \\ & & + 
  \frac{5034016 \eta_{4,10} \mu_{3,3}}{59049} + 
  \frac{5680 \eta_{0,1} \mu_{5,1}}{6561} - 
  \frac{5680 \eta_{2,1} \mu_{5,1}}{59049} + 
  \frac{152960 \eta_{2,2} \mu_{5,1}}{177147}\\ & & - 
 \frac{369200 \eta_{2,3} \mu_{5,1}}{177147} + 
  \frac{26240 \eta_{4,1} \mu_{5,1}}{4782969} - 
  \frac{717680 \eta_{4,2} \mu_{5,1}}{1594323} + 
  \frac{227200 \eta_{4,3} \mu_{5,1}}{531441} \\ & &- 
  \frac{1804544 \eta_{4,4} \mu_{5,1}}{4782969} - 
  \frac{803200 \eta_{4,5} \mu_{5,1}}{1594323} + 
  \frac{738400 \eta_{4,6} \mu_{5,1}}{177147} + 
  \frac{369200 \eta_{4,8} \mu_{5,1}}{1594323}\\ & & - 
  \frac{9942400 \eta_{4,9} \mu_{5,1}}{4782969} + 
  \frac{40021280 \eta_{4,10} \mu_{5,1}}{4782969} + 
  \frac{304 \eta_{0,1} \mu_{5,2}}{2187} - 
  \frac{105136 \eta_{2,1} \mu_{5,2}}{177147}\\ & & - 
  \frac{103040 \eta_{2,2} \mu_{5,2}}{59049} - 
  \frac{19760 \eta_{2,3} \mu_{5,2}}{59049} - 
  \frac{730240 \eta_{4,1} \mu_{5,2}}{4782969} - 
  \frac{1423184 \eta_{4,2} \mu_{5,2}}{1594323}\\ & & - 
  \frac{352640 \eta_{4,3} \mu_{5,2}}{531441} + 
  \frac{91904 \eta_{4,4} \mu_{5,2}}{1594323} + 
  \frac{7936640 \eta_{4,5} \mu_{5,2}}{4782969} + 
  \frac{39520 \eta_{4,6} \mu_{5,2}}{59049}\\ & & + 
  \frac{6833840 \eta_{4,8} \mu_{5,2}}{4782969} + 
  \frac{6697600 \eta_{4,9} \mu_{5,2}}{1594323} + 
  \frac{2141984 \eta_{4,10} \mu_{5,2}}{1594323} + 
  \frac{80 \eta_{0,1} \mu_{5,3}}{81}\\ & & - 
\frac{80 \eta_{2,1} \mu_{5,3}}{729} + 
  \frac{26240 \eta_{2,2} \mu_{5,3}}{177147} - 
  \frac{5200 \eta_{2,3} \mu_{5,3}}{2187} + 
  \frac{390272 \eta_{4,1} \mu_{5,3}}{1594323}\\ & & - 
  \frac{1519120 \eta_{4,2} \mu_{5,3}}{1594323} + 
  \frac{3200 \eta_{4,3} \mu_{5,3}}{6561} + 
  \frac{84736 \eta_{4,4} \mu_{5,3}}{1594323} + 
  \frac{18560 \eta_{4,5} \mu_{5,3}}{59049}\\ & & + 
  \frac{10400 \eta_{4,6} \mu_{5,3}}{2187} + 
  \frac{5200 \eta_{4,8} \mu_{5,3}}{19683} - 
  \frac{1705600 \eta_{4,9} \mu_{5,3}}{4782969} + 
  \frac{563680 \eta_{4,10} \mu_{5,3}}{59049}\\ & & + 
  \frac{1280 \eta_{0,1} \mu_{5,4}}{2187} + 
  \frac{47360 \eta_{2,1} \mu_{5,4}}{59049} + 
  \frac{40960 \eta_{2,2} \mu_{5,4}}{177147} - 
  \frac{83200 \eta_{2,3} \mu_{5,4}}{59049}\\ & & + 
  \frac{9060352 \eta_{4,1} \mu_{5,4}}{4782969} - 
  \frac{98560 \eta_{4,2} \mu_{5,4}}{531441} - 
  \frac{409600 \eta_{4,3} \mu_{5,4}}{1594323} + 
  \frac{212992 \eta_{4,4} \mu_{5,4}}{4782969}\\ & & + 
  \frac{1515520 \eta_{4,5} \mu_{5,4}}{4782969} + 
  \frac{166400 \eta_{4,6} \mu_{5,4}}{59049} - 
  \frac{3078400 \eta_{4,8} \mu_{5,4}}{1594323} - 
  \frac{2662400 \eta_{4,9} \mu_{5,4}}{4782969}\\ & & + 
  \frac{9018880 \eta_{4,10} \mu_{5,4}}{1594323} + 
  \frac{5632 \eta_{0,1} \mu_{5,5}}{6561} + 
  \frac{15872 \eta_{2,1} \mu_{5,5}}{19683} - 
  \frac{212992 \eta_{2,2} \mu_{5,5}}{177147}\\ & & - 
  \frac{366080 \eta_{2,3} \mu_{5,5}}{177147} - 
  \frac{163840 \eta_{4,1} \mu_{5,5}}{531441} - 
  \frac{1235456 \eta_{4,2} \mu_{5,5}}{4782969} - 
  \frac{229376 \eta_{4,3} \mu_{5,5}}{1594323}\\ & & + 
 \frac{360448 \eta_{4,4} \mu_{5,5}}{531441} - 
  \frac{409600 \eta_{4,5} \mu_{5,5}}{531441} + 
  \frac{732160 \eta_{4,6} \mu_{5,5}}{177147} - 
  \frac{1031680 \eta_{4,8} \mu_{5,5}}{531441}\\ & & + 
  \frac{13844480 \eta_{4,9} \mu_{5,5}}{4782969} + 
  \frac{39683072 \eta_{4,10} \mu_{5,5}}{4782969} - 
  \frac{2080 \eta_{0,1} \mu_{5,6}}{243} + 
  \frac{2080 \eta_{2,1} \mu_{5,6}}{2187}\\ & & + 
  \frac{116480 \eta_{2,2} \mu_{5,6}}{6561} + 
  \frac{3117920 \eta_{2,3} \mu_{5,6}}{177147} - 
  \frac{83200 \eta_{4,1} \mu_{5,6}}{59049} + 
  \frac{713440 \eta_{4,2} \mu_{5,6}}{177147}\\ & & - 
  \frac{83200 \eta_{4,3} \mu_{5,6}}{19683} - 
  \frac{2030080 \eta_{4,4} \mu_{5,6}}{177147} - 
  \frac{1015040 \eta_{4,5} \mu_{5,6}}{177147} - 
  \frac{226516160 \eta_{4,6} \mu_{5,6}}{1594323}\\ & & - 
  \frac{17039360 \eta_{4,7} \mu_{5,6}}{531441} - 
  \frac{3117920 \eta_{4,8} \mu_{5,6}}{1594323} - 
  \frac{174603520 \eta_{4,9} \mu_{5,6}}{4782969} - 
  \frac{604222528 \eta_{4,10} \mu_{5,6}}{4782969}\\ & & + 
  \frac{4259840 \eta_{2,3} \mu_{5,7}}{177147} + 
  \frac{42598400 \eta_{4,6} \mu_{5,7}}{531441} - 
  \frac{54525952 \eta_{4,7} \mu_{5,7}}{4782969} - 
  \frac{80936960 \eta_{4,8} \mu_{5,7}}{4782969}\\ & & - 
  \frac{34078720 \eta_{4,9} \mu_{5,7}}{4782969} - 
  \frac{417464320 \eta_{4,10} \mu_{5,7}}{4782969} + 
  \frac{5200 \eta_{0,1} \mu_{5,8}}{2187} - 
  \frac{5200 \eta_{2,1} \mu_{5,8}}{19683}\\ & & + 
  \frac{457600 \eta_{2,2} \mu_{5,8}}{177147} - 
  \frac{604240 \eta_{2,3} \mu_{5,8}}{59049} + 
  \frac{10391680 \eta_{4,1} \mu_{5,8}}{4782969} - 
  \frac{494000 \eta_{4,2} \mu_{5,8}}{177147}\\ & & + 
  \frac{208000 \eta_{4,3} \mu_{5,8}}{177147} - 
  \frac{615680 \eta_{4,4} \mu_{5,8}}{531441} - 
  \frac{7030400 \eta_{4,5} \mu_{5,8}}{4782969} + 
  \frac{15589600 \eta_{4,6} \mu_{5,8}}{1594323}\\ & & + 
  \frac{28966912 \eta_{4,7} \mu_{5,8}}{4782969} + 
  \frac{604240 \eta_{4,8} \mu_{5,8}}{531441} - 
  \frac{53173120 \eta_{4,9} \mu_{5,8}}{4782969} + 
  \frac{25170080 \eta_{4,10} \mu_{5,8}}{531441}\\ & & + 
  \frac{16640 \eta_{0,1} \mu_{5,9}}{6561} + 
  \frac{615680 \eta_{2,1} \mu_{5,9}}{177147} - 
  \frac{532480 \eta_{2,2} \mu_{5,9}}{177147} - 
  \frac{1933568 \eta_{2,3} \mu_{5,9}}{177147}\\ & & + 
  \frac{7454720 \eta_{4,1} \mu_{5,9}}{4782969} + 
  \frac{416000 \eta_{4,2} \mu_{5,9}}{4782969} - 
  \frac{5324800 \eta_{4,3} \mu_{5,9}}{4782969} + 
  \frac{7454720 \eta_{4,4} \mu_{5,9}}{4782969}\\ & & - 
  \frac{19701760 \eta_{4,5} \mu_{5,9}}{4782969} + 
  \frac{49886720 \eta_{4,6} \mu_{5,9}}{4782969} - 
  \frac{109051904 \eta_{4,7} \mu_{5,9}}{4782969} - 
  \frac{71542016 \eta_{4,8} \mu_{5,9}}{4782969}\\ & & + 
  \frac{61874176 \eta_{4,9} \mu_{5,9}}{4782969} + 
  \frac{80544256 \eta_{4,10} \mu_{5,9}}{1594323} - 
  \frac{112736 \eta_{0,1} \mu_{5,10}}{6561} + 
  \frac{112736 \eta_{2,1} \mu_{5,10}}{59049}\\ & & + 
 \frac{6313216 \eta_{2,2} \mu_{5,10}}{177147} + 
  \frac{5034016 \eta_{2,3} \mu_{5,10}}{59049} - 
  \frac{4509440 \eta_{4,1} \mu_{5,10}}{1594323} + 
  \frac{38668448 \eta_{4,2} \mu_{5,10}}{4782969}\\ & & - 
  \frac{4509440 \eta_{4,3} \mu_{5,10}}{531441} - 
  \frac{110030336 \eta_{4,4} \mu_{5,10}}{4782969} - 
  \frac{55015168 \eta_{4,5} \mu_{5,10}}{4782969} - 
  \frac{604222528 \eta_{4,6} \mu_{5,10}}{4782969}\\ & & + 
  \frac{166985728 \eta_{4,7} \mu_{5,10}}{4782969} - 
  \frac{5034016 \eta_{4,8} \mu_{5,10}}{531441} - 
  \frac{281904896 \eta_{4,9} \mu_{5,10}}{1594323} - 
  \frac{279502912 \eta_{4,10} \mu_{5,10}}{531441}
\end{eqnarray*}
\bibliographystyle{JHEP}

\end{document}